\documentclass[twocolumn,showpacs,preprintnumbers,amsmath,amssymb]{revtex4}%
\usepackage[dvips]{graphicx}
\usepackage{graphics}
\usepackage{dcolumn}
\usepackage{bm}
\usepackage{amsmath}
\usepackage{amsfonts}
\usepackage{amssymb}%
\begin{document}
\title[Conductivity of Quantized Multilayer Films]{ Quantum Size Effect in Conductivity of Multilayer Metal Films}
\author{A. E. Meyerovich}
\affiliation{Kingston, RI 02881-0817 }
\author{I. V. Ponomarev}
\email{ilya@qc.physics.edu}
\affiliation{Flushing, NY 11367}
\date{\today }

\begin{abstract}
Conductivity of quantized multilayer metal films is analyzed with an emphasis
on scattering by rough interlayer interfaces. Three different types of quantum
size effect (QSE) in conductivity are predicted. Two of these QSE are similar
to those in films with scattering by rough walls. The third type of QSE is
unique and is observed only for certain positions of the interface. The
corresponding peaks in conductivity are very narrow and high with a finite
cutoff which is due only to some other scattering mechanism or the smearing of
the interface. There are two classes of these geometric resonances. Some of
the resonance positions of the interface are universal and do not depend on
the strength of the interface potential while the others are sensitive to this
potential. This geometric QSE gradually disappears with an increase in the
width of the interlayer potential barrier.

\end{abstract}
\pacs{73.40.-c, 73.63.Hs, 72.10.Fk, 73.50.-h}
\maketitle




\section{Introduction}

Boundary scattering is essential for complete description of nanosystems such
as quantum wells, ultrathin films or wires, \emph{etc.} Due to the large
surface-to-volume ratio, boundaries are expected to play a much greater role
in determining the overall properties in a nanostructure than in a bulk
material. For example, recent scanning tunneling microscopy (STM) data have
shown that electron energy spectra can be more strongly correlated to the
buried interfacial lattices than to the surface immediately beneath the STM
tip \cite{alt1}. These observations clearly indicate that a small lateral
variation along the boundary can have a significant long range effect in a
semi-ballistic electron system. Thus, a more realistic description of a
nanoscale quantized system must go beyond the common perfect geometric
boundary and include boundary corrugations. Indeed, random surface roughness
of a thin metal film can dominate incoherent scattering and relaxation, and
can lead to anomalous quantum size effect such as large oscillatory dependence
of the in-plane conductivity on the film thickness \cite{pon1}.

The same must be true not only for the quantum well (film) walls but also for
the interlayer interfaces in multilayer films. It is well known that the
roughness of the interlayer interfaces plays an important role in, for
example, giant magnetoresistance (see review \cite{levy1} and references
therein). The purpose of this paper is to analyze the effect of irregular
corrugation of the interlayer interfaces on the lateral conductivity of
quantized multilayer films without magnetic effects. We will see that the
interface scattering can result in unique features of the quantum size effect
(QSE) which are strikingly different from QSE with scattering by bulk or wall
inhomogeneities. Orbital and spin magnetic effects of the type studied in
Refs. \cite{joy} will be studied separately.

In ultrathin films, the motion of electrons across the films can be quantized.
QSE in metal films is studied experimentally by measuring conductivity
\cite{conduct1,qse2} and susceptibility \cite{susc1} of the films or in
spectroscopy \cite{spect1} and STM \cite{alt1} measurements (for earlier
results, see references therein). As a result of QSE, the $3D$ electron
spectrum $\epsilon\left(  \mathbf{p}\right)  $ splits into a set of minibands
$\epsilon_{j}\left(  \mathbf{q}\right)  $ where $\mathbf{q}$ is the $2D$
momentum along the film ($yz-$plane). In the simplest case of a single-layer
film approximated by a rectangular quantum well, the quantized values of the
$x-$component of momentum are $p_{x_{j}}=\pi j/L$ (here and below $\hbar=1$).
If in such quantized metal films the Fermi energy $E_{F}$ is unaffected by the
quantization, the Fermi surface reduces to a set of $2D$ curves $\epsilon
_{Fj}\left(  \mathbf{q}\right)  $ that correspond to cross-sections of the
$3D$ Fermi surface $\epsilon\left(  \mathbf{p}\right)  =E_{F}$ by a set of
planes $p_{x_{j}}=\pi j/L$, $\epsilon_{Fj}\left(  \mathbf{q}\right)
=\epsilon_{F}\left(  p_{x_{j}},\mathbf{q}\right)  $.

This quantization of motion, which is determined by the film thickness $L$,
leads to several types of QSE. First, any change of the film thickness $L$
results in the change in size and number of the Fermi curves $\epsilon
_{Fj}\left(  \mathbf{q}\right)  $. This thickness-driven change in number of
the Fermi curves $\epsilon_{Fj}\left(  \mathbf{q}\right)  $\ [or, what is the
same, number of occupied minibands $\epsilon_{j}\left(  \mathbf{q}\right)  $]
leads to a singularity in the density of states. These singularities are the
most obvious manifestations of QSE.

These singularities in the density of states, by itself, do not lead to any
\emph{striking} anomalies in the dependence of the lateral conductivity
$\sigma$ of the film on the thickness $L$. The conductivity is more sensitive
to electron scattering than to the density of states. However, the change in
the number of the occupied minibands $S$ can be accompanied by a change in
number of allowed scattering channels that correspond to the scattering-driven
electron transitions between minibands $\epsilon_{j}\left(  \mathbf{q}\right)
$. The effect of this step-like change in the number of scattering channels on
the conductivity is much stronger than that of the singularities in the
density of states \cite{ser1}. When all scattering-driven interband
transitions are allowed, QSE manifests itself as a pronounced saw-like
dependence of the conductivity on the film thickness. This type of QSE in
quantized films has been predicted both for scattering by impurities and
surface inhomogeneities \cite{sand, ash}.

When the main scattering mechanism is the scattering by surface
inhomogeneities, many of the interband transitions can often be suppressed.
This happens, for example, when the average size of the surface
inhomogeneities $R$ is much larger than the the thickness of the film and/or
the particle wavelength $\lambda_{F}$. Then the usual QSE, which is described
above, disappears and is replaced by a different kind of the size effect
\cite{pon1}. This anomalous QSE, which is somewhat reminiscent of the magnetic
breakthrough, is completely decoupled from the singularities in the density of
states and is associated solely with opening of interband scattering channels
for gliding electrons at certain values of the film thickness, $L_{i}%
\simeq\sqrt{\left(  i+1/2\right)  R\lambda_{F}/2}$.

The main goal of this paper is to analyze QSE in \emph{multilayer} films with
an emphasis on the scattering by the interface between the layers. We will see
that, in addition to the above two types of QSE, the multilayer films can
exhibit a peculiar "geometric" QSE with very narrow high peaks in the lateral
conductivity. Some of the positions of these spikes in conductivity are
universal; these spikes appear when the ratio of the thicknesses of the film
layers is given by simple fractions. The position of the rest of the spikes
depend on the strength of the interlayer interface.

In the next Section, we briefly present the main equations for the
conductivity and introduce proper dimensionless variables. The results are
presented in Section III. Section IV contains the summary and brief discussion
of the results. Appendix contains auxiliary information on the energy spectrum
of multilayer films of the type used in the calculations.

\section{Conductivity}

\subsection{Scattering by the interlayer interface}

For simplicity, we consider an ultrathin film of thickness $L$ consisting of
only two layers with the thickness of $L_{1}$\ and\ $L_{2}$. The interface
between the layers is rough with random corrugation. The exact position of the
interface, $x=L_{1}+\xi\left(  y,z\right)  ,$ is described by the random
function $\xi\left(  y,z\right)  $ with the zero average, $\left\langle
\xi\right\rangle =0$. The random interface inhomogeneities $\xi\left(
y,z\right)  $ are best characterized by the correlation function $\zeta\left(
\mathbf{s}\right)  $,
\begin{align}
\zeta\left(  \mathbf{s}\right)   & = \zeta\left(  \left\vert \mathbf{s}%
\right\vert \right) =\left\langle \xi(\mathbf{s}_{1})\xi(\mathbf{s}%
_{1}+\mathbf{s})\right\rangle \nonumber\\
& \equiv A^{-1}\int\xi(\mathbf{s}_{1})\xi(\mathbf{s}_{1}+\mathbf{s}%
)d\mathbf{s}_{1}, \label{aa1}%
\end{align}
where the vector $\mathbf{s}$ gives the $2D$ coordinates along the interface
and $A$ is the averaging area. Here, it is\ assumed that the correlation
properties of the surface do not depend on direction. Two main characteristics
of the surface correlation functions $\zeta$ are the average amplitude
(\textquotedblright height\textquotedblright) and the correlation radius
(\textquotedblright size\textquotedblright) of surface inhomogeneities, $\ell$
and $R$.

To emphasize the scattering by inhomogeneities of the interlayer interface, we
start from films with ideal outside walls that do not contribute to electron
scattering. The combined effect of interface and wall inhomogeneities will be
considered elsewhere.

Mostly we are interested in the dependence of the lateral conductivity on the
film thickness and have in mind the following experimental situation. The
first layer of the film is grown on some (ideal) substrate. The surface is
then roughened by adding inhomogeneous adsorbate or by some other means. The
growth of the second layer starts from this rough interface and the
conductivity is measured at different values of $L_{2}$ either in the process
of growth or after the growth is completed. An advantage of such setup with a
buried interface is that it allows one to measure the conductivity at various
values of the film thickness with \emph{exactly the same} random rough interface.

In this setup, the thickness of the first layer, $L_{1}$, should be considered
as fixed, while the thickness of the second layer, $L_{2}$, is variable. Below
we are calculate the film conductivity $\sigma$\ as a function of the film
thickness $L=L_{1}+L_{2}$, $\sigma\left(  L\right)  $, assuming that
$L_{1}=const$. The measurements of conductivity can be performed in stationary
conditions at different values of $L_{2}$ or as a function of time, in the
process of film growth as in Ref. \cite{pal1}.

The second layer can be made of the same or different material as the first.
If the material is different, then the electron potential energy between the
layers differs by some $\Delta U$. The the structure of the energy spectrum
becomes a complicated function of $\Delta U$ making the behavior of
conductivity highly irregular \cite{arm2}.

Below we consider both layers to be made of the same material with the
interface being the only disruption in the potential relief. Then the simplest
model of the interface is the $\delta-$functional potential barrier%
\begin{equation}
U=U_{0}\delta\left(  x-L_{1}-\xi\left(  y,z\right)  \right)  . \label{b1}%
\end{equation}
This immediately introduces two new physical parameters into the problem: the
strength of the barrier $U_{0}$ and its (average) position $L_{1}$. In what
follows, we study the dependence of the conductivity on these parameters. When
necessary, instead of the $\delta$-function we will study the corrugated
interface with the finite width $D$. In experiment, the barrier can be a
dislocation wall, twin boundary, or an oxide or dielectric layer (see,
\emph{e.g., }Ref. \cite{bul1} and references therein).

The presence of the interface $\left(  \ref{b1}\right)  $ changes the
spectrum. When calculating the changes in the spectrum, one can ignore small
corrugation $\xi\left(  y,z\right)  $. The changes in spectrum caused by the
$\delta-$type barrier $\left(  \ref{b1}\right)  $ are discussed in Appendix.
The random corrugation of the interface is responsible for the electron
scattering and gives rise to the collision operator in the transport equation.

The scattering by the interface inhomogeneities leads to the transitions
between the states $\epsilon_{i}\left(  \mathbf{q}\right)  \rightarrow
\epsilon_{j}\left(  \mathbf{q}^{\prime}\right)  $. Several ways of calculating
the corrugation-driven transition probabilities $W_{ij}\left(  \mathbf{q,q}%
^{\prime}\right)  $ are described in Ref. \cite{arm2}. The simplest methods
are either the direct perturbation approach \cite{fish1} or the mapping
transformation method \cite{map1} with both giving the same result in most of
the parameter range.

The corrugation-driven contribution $\delta U$ to the interface potential, Eq.
$\left(  \ref{b1}\right)  $, with small corrugation $\xi$ is%
\begin{equation}
\delta U=-U_{0}\xi\left(  y,z\right)  \delta^{\prime}\left(  x-L_{1}\right)  .
\label{b2}%
\end{equation}
The matrix element $V_{ij}\left(  \mathbf{q,q}^{\prime}\right)  $ of this
perturbation between the states $\epsilon_{j}\left(  \mathbf{q}\right)
,\epsilon_{j}\left(  \mathbf{q}^{\prime}\right)  $ is
\begin{align}
&  V_{ij}=-U_{0}\int e^{i\mathbf{s}( \mathbf{q-q}^{\prime})} \xi\left(
\mathbf{s}\right)  \Psi_{i}\left(  x\right)  \delta^{\prime}\left(
x-L_{1}\right)  \Psi_{j}\left(  x\right)  dxd\mathbf{s}\nonumber\\
& =U_{0}\xi\left(  \mathbf{q-q}^{\prime}\right)  \left[  \Psi_{i}\left(
L_{1}\right)  \Psi_{j}^{\prime}\left(  L_{1}\right)  +\Psi_{i}^{\prime}\left(
L_{1}\right)  \Psi_{j}\left(  L_{1}\right)  \right]  ,\label{b3}%
\end{align}
where $\Psi_{i}\left(  x\right)  $ are the quantized wave functions for
electron motion across the film. Note that the derivatives $\Psi^{\prime
}\left(  x\right)  $ for films with a $\delta-$type barrier inside are
discontinuous at the position of the barrier $x=L_{1}$. Therefore, $\Psi
_{i}^{\prime}\left(  L_{1}\right)  $ in Eq. $\left(  \ref{b3}\right)  $ should
be understood as $\Psi_{i}^{\prime}\left(  L_{1}\right)  =\left[  \Psi
_{i}^{\prime}\left(  L_{1}+0\right)  +\Psi_{i}^{\prime}\left(  L_{1}-0\right)
\right]  /2$.

The corrugation-driven transition probability $W_{ij}\left(  \mathbf{q}%
,\mathbf{q}^{\prime}\right)  $ is given by the square of this matrix element
which should be averaged over the random inhomogeneities $\xi$:
\begin{align}
W_{ij}\left(  \mathbf{q},\mathbf{q}^{\prime}\right)   &  =\left\langle
\left\vert V_{ij}\left(  \mathbf{q,q}^{\prime}\right)  \right\vert
^{2}\right\rangle _{\xi}=U_{0}^{2}\zeta\left(  \left\vert \mathbf{q}%
_{i}\mathbf{-q}_{j}^{\prime}\right\vert \right)  G_{ij},\label{b4}\\
G_{ij}  &  =\left[  \Psi_{i}\left(  L_{1}\right)  \Psi_{j}^{\prime}\left(
L_{1}\right)  +\Psi_{i}^{\prime}\left(  L_{1}\right)  \Psi_{j}\left(
L_{1}\right)  \right]  ^{2}, \label{b5}%
\end{align}
where $\zeta\left(  \left\vert \mathbf{q}_{i}\mathbf{-q}_{j}^{\prime
}\right\vert \right)  $ is the Fourier image of the correlation function of
the interface inhomogeneities $\left(  \ref{aa1}\right)  $. The coefficients
$G_{ik}$ are calculated with the help of the wave functions presented in the
Appendix. The explicit form of $G_{ik}$ is given in the next subsection.

The transport equation is a set of equations for the electron distribution
functions $n_{i}\left(  \mathbf{q}\right)  $ in minibands $\epsilon_{i}$ and
has the standard Boltzmann-Waldmann-Snider form \cite{arm2}:
\begin{equation}
\frac{dn_{i}}{dt}=2\pi\sum_{j}\int W_{ij}\left[  n_{j}-n_{i}\right]
\delta\left(  \epsilon_{i\mathbf{q}}-\epsilon_{j\mathbf{q}^{\prime}}\right)
\frac{d^{2}q^{\prime}}{\left(  2\pi\right)  ^{2}}. \label{e6}%
\end{equation}

The integration over $dq^{\prime}$ is done using the $\delta$-function,
$\delta\left(  \epsilon_{i\mathbf{q}}-\epsilon_{j\mathbf{q}^{\prime}}\right)
=m_{ij}^{\ast}\delta\left(  q^{\prime}-q_{ij}\right)  /q_{ij}$, where
$q_{ij}\left(  q\right)  $ is the solution of the equation $\epsilon
_{j}\left(  \mathbf{q}_{ij}\right)  =\epsilon_{i}\left(  \mathbf{q}\right)  $
and the effective masses $m_{ij}^{\ast}=q_{ij}/\left(  \partial\epsilon
_{j}/\partial q\right)  \left\vert _{q=q_{ij}}\right.  $. As always in the
transport theory, the angular integration is eliminated by using the angular
harmonics. The current is given by the first harmonic of the distribution
$n_{i}^{\left(  1\right)  }\equiv\nu_{i}$ the equation for which involves only
the zeroth and first harmonics $W_{ij}^{\left(  0,1\right)  }\left(
q,q_{ij}\right)  $ of $W\left(  \mathbf{q-q}_{ij}\right)  $ over the angle
$\widehat{\mathbf{qq}}_{jj^{\prime}}$,
\begin{align}
d\nu_{i}\left(  q\right)  /dt  &  =-\sum_{j}\nu_{j}\left(  q_{ij}\right)
/\tau_{ij},\label{ee6}\\
\frac{1}{\tau_{ij}}  &  =\frac{m}{2}\sum_{k}\left[  \delta_{ij}W_{ik}^{\left(
0\right)  }-\delta_{jk}W_{ij}^{\left(  1\right)  }\right]  ,\nonumber\\
W_{ij}^{\left(  0,1\right)  }  &  =U_{0}^{2}\zeta^{\left(  0,1\right)
}\left(  \mathbf{q}_{i}-\mathbf{q}_{j}\right)  G_{ij},\nonumber
\end{align}
where we, to simplify the equations, assume that the effective mass
$m_{ij}^{\ast}$ does not depend on its indices, $m=m_{ij}^{\ast}$.

The solution of Eqs. $\left(  \text{\ref{ee6}}\right)  $ provides the
$2D$\ conductivity of the film:
\begin{equation}
\sigma=-\frac{e^{2}}{3\hbar^{2}}\sum_{i}\nu_{i}\left(  q_{i}\right)  q_{i}.
\label{ee7}%
\end{equation}

\subsection{Dimensionless variables}

The problem involves several length scales - particle Fermi wavelength
$\lambda_{F}=\pi/p_{F},$ the thickness of the layers $L_{1}$ and $L_{2}%
$\ ($L_{1}+L_{2}=L$), the correlation radius of the surface inhomogeneities
$R$, and the interface thickness $D$. Another length parameter, the amplitude
of inhomogeneities $\ell$, is perturbative and enters conductivity as a
coefficient,
\begin{equation}
\sigma=\frac{2e^{2}}{\hbar}\frac{R^{2}}{\ell^{2}}f\left(  \lambda_{F}%
,L_{i},R,D\right)  . \label{c2}%
\end{equation}
Note, that we consider only the contribution from surface roughness and
disregard the bulk scattering. As a result, the conductivity $\left(
\text{\ref{c2}}\right)  $ diverges in the limit of vanishing inhomogeneities
$\ell\rightarrow0$\ or $R\rightarrow\infty$. The proper account of bulk
scattering \cite{arm3} eliminates this divergence.

It is convenient to measure all the length parameters in the units of the
Fermi wavelength $\lambda_{F}=\pi/p_{F}$. Instead of the interface strength
$U_{0},$ we use interchangeably two equivalent dimensionless parameters $g$
and $u_{0}$,%
\begin{equation}
g=u_{0}L/\pi\lambda_{F}=2mU_{0}\lambda_{F}L/\pi\hbar^{2} \label{b6}%
\end{equation}
($g$ is convenient for the calculation of the spectrum while $u_{0}$ is a
proper energy parameter for characterization of the conductivity in our
setup). The position of the interface is characterized by the parameter
$\delta,$%
\begin{equation}
\delta=L_{2}/L. \label{b7}%
\end{equation}
In computations, $\delta$ changes from $0$ (no second layer) to $1$ (the
second layer much wider than the first). It is worth repeating that we are
looking at the experimental situation when the thickness of the first layer is
fixed and the conductivity is measured as a function of the thickness of the
second layer (or the overall film thickness).

The energy spectrum $\epsilon_{i}\left(  \mathbf{q}\right)  $ is described by
dimensionless energy units $z_{i},$%
\begin{equation}
\epsilon_{i}\left(  \mathbf{q}\right)  =\frac{1}{2m}\left(  \frac{\pi^{2}%
}{L^{2}}z_{i}^{2}+q^{2}\right)  , \label{b8}%
\end{equation}
where $z_{i}$ is given by the solution of the $1D$ Schroedinger equation for a
quantum well with a $\delta-$type barrier inside (see Appendix):%

\begin{equation}
\sin\pi z+\frac{g}{z}\sin\left(  \pi z\delta\right)  \sin\left[  \pi z\left(
1-\delta\right)  \right]  =0, \label{b9}%
\end{equation}

Finally, the conductivity $\sigma\left(  L\right)  $ for the experimental
setup which has been described above, will be displayed by the dimensionless
function $f_{L}\left(  L/\lambda_{F}\right)  ,$%
\begin{equation}
\sigma\left(  L\right)  =\frac{2e^{2}}{\hbar}\frac{R^{2}}{\ell^{2}}f_{L},
\label{c3}%
\end{equation}
for various values of $R/\lambda_{F}$, $D/\lambda_{F}$, $L_{1}/\lambda_{F}$
and the strength of the barrier $u_{0}$.

All the figures below present this dimensionless function $f_{L}$. This
function is plotted under the assumption that the experiment is performed at
fixed thickness of the first layer. For uniformity, the figures for weak
interfaces are plotted for $u_{0}=0.1$, and for strong interface barriers -
for $u_{0}=10$. The simplest energy spectrum corresponds to thin first layers,
$\lambda_{F}\leq L_{1}<2\lambda_{F}$. Therefore, for transparency of results,
the majority of the data are presented for $L_{1}/\lambda_{F}=1.1$ (for
comparison, some of the graphs give the conductivity for larger $L_{1}$).

The computational results below are presented for the Gaussian correlation
function of the interface inhomogeneities,%
\begin{equation}
\zeta\left(  \mathbf{s}\right)  =\ell^{2}\exp\left(  -s^{2}/2R^{2}\right)  .
\label{b21}%
\end{equation}
The angular harmonics for this correlator, which enter the transition
probabilities in Eq. $\left(  \text{\ref{ee6}}\right)  $, are equal to
\begin{align}
\zeta^{\left(  0\right)  }\left(  q_{i},q_{j}\right)   &  =4\pi\ell^{2}%
R^{2}\left[  e^{-QQ^{\prime}}I_{0}\left(  QQ^{\prime}\right)  \right]
e^{-\left(  Q-Q^{\prime}\right)  ^{2}/2},\label{b22}\\
\zeta^{\left(  1\right)  }\left(  q_{i},q_{j}\right)   &  =4\pi\ell^{2}%
R^{2}\left[  e^{-QQ^{\prime}}I_{1}\left(  QQ^{\prime}\right)  \right]
e^{-\left(  Q-Q^{\prime}\right)  ^{2}/2}\nonumber
\end{align}
where $Q=q_{i}R$, $Q^{\prime}=q_{j}R$.

Analysis of QSE in Ref. \cite{pon1} for ultrathin films with scattering by the
film walls demonstrated that the results for all types of correlators are
qualitatively the same as for the Gaussian one when $R\ll L$. For large
inhomogeneities, $R\gg L$, the results for all types of correlators with
exponential power spectra are similar to those for the Gaussian correlator and
are qualitatively different from the power-law correlators. The results for
the power-law correlators are less interesting: such films always exhibit the
standard saw-like QSE irrespective of the value of $R$ because of the wider
fluctuations of the inhomogeneity sizes. Therefore, in this paper we consider
only the exponential correlators with a well-defined size of inhomogeneities.

\section{Results}

\subsection{Standard quantum size effect}

The standard quantum size effect (QSE) in films manifests itself by a saw-like
dependence of the conductivity $\sigma$ on the film thickness $L$
\cite{sand,ash}. The positions of the singularities - the saw teeth -
correspond to the values of thickness at which a new energy miniband
$\epsilon_{j}$ becomes accessible. The amplitude of the conductivity drop in
such a singular point depends, in the case of scattering by surface
inhomogeneities, on the effectiveness of the roughness-driven interband
transitions. If the probability of such transitions $W_{i\neq j}$, is small in
comparison to the rate of the intraband scattering $W_{ii}$, the singularities
in the curves $\sigma\left(  L\right)  $ are almost completely suppressed and
the standard QSE disappears \cite{ser1}.

Analysis of the roughness-driven transition probabilities for surface
scattering in Ref.\cite{pon1} for different classes of surface roughness
showed that, when the average size of inhomogeneities, $R$, is much smaller
than the film thickness $L$, the values of the interband transition
probabilities $W_{i\neq j}$ are comparable to that for the intraband
scattering $W_{ii}$, all the scattering channels are equally important. In
this case, the curves $\sigma\left(  L\right)  $ always exhibit the standard
QSE. The same should be true for scattering by the interlayer interfaces. This
is illustrated in Figures 1 and 2 which show $\sigma\left(  L\right)  $ for a
weak \ and strong interface potentials $u_{0}=0.1$ and $u_{0}=10$
respectively. In Figure 1 the thickness of the first layer is $L_{1}%
=2.1\ \lambda_{F}$, in Figure 2 - $L_{1}=1.1\ \lambda_{F}$. In both figures,
the size of inhomogeneities is $R=\lambda_{F}$. Both figures exhibit a
well-pronounced saw-like structure. The positions of the singularities for the
weak interface are almost equidistant reflecting the fact the energy structure
is close to that for a square well without perturbation inside. The strong
interface affects the energy spectrum and, therefore, the positions and the
shapes of the saw teeth. However, at very large film thickness $L\gg L_{1}$
the interface is located very close to the well wall and the spectrum start to
recover its unperturbed structure. This manifests itself in a recovery of the
equidistant distribution of the singularities in Figure 2 at large $L$.
Because of a peculiar dependence of the transition probabilities on the
interface strength (see Appendix), the conductivity grows much faster with
increasing film thickness in the case of the weak interface than for the
strong interface.%

\begin{figure}[ptb]
\includegraphics{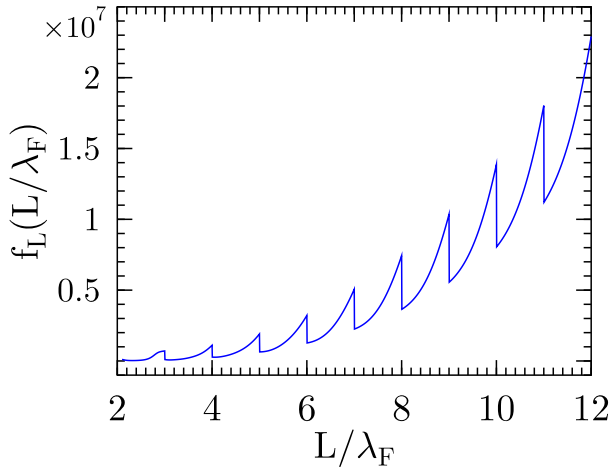}
\caption
{Dimensionless conductivity of quantized films, Eq. (15), as a function of the film thickness $L$.
The saw-like dependence is typical for the standard quantum size effect. The correlation
radius of inhomogeneities $R/ \lambda_F =1$,
the thickness of the first layer $L_1/ \lambda
_F =2.1$, the width of the interface
$d=D/ \lambda_F =0.01$, and the strength of the barrier $u_0=0.1$.}
\label{fig1}
\end{figure}%
%

\begin{figure}[ptb]
\includegraphics{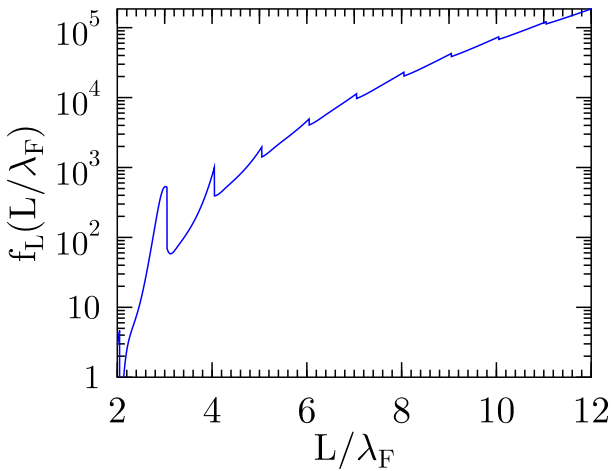}
\caption
{Standard QSE in conductivity of quantized films, Eq. (15), as a function of the film thickness $L$
for strong interface potential, $u_0=10$. The correlation radius of inhomogeneities $R/ \lambda
_F =1$,
the thickness of the first layer $L_1/ \lambda
_F =1.1$, the width of the interface
$d=10^{-4}$. }
\label{fig2}
\end{figure}%

\subsection{Quantum size effect for large-scale inhomogeneities}

The standard QSE of the type described in the previous subsection disappears
in the single-layer film when the correlation size of inhomogeneities, $R$, is
larger than the film thickness, $R\gg L$, \emph{and }the correlation function
in the momentum space $\zeta\left(  \mathbf{q}\right)  $ (the so-called power
spectrum of inhomogeneities) decays exponentially at large wave numbers
$\mathbf{q}$. Instead, the single-layer films exhibit an anomalous QSE
\cite{pon1}.

The explanation involves the interband transitions. It seems that at large $R$
the off-diagonal $W_{i\neq k}$ are small and the interband transitions are
suppressed. However, at\ certain values of large $L$, few of the elements
$W_{i\neq k}$, which are close to the main diagonal, could become comparable
to $W_{ii}$ even for large $R$. Then the transitions $i\leftrightarrow i+1$
could become noticeable leading to a drop in conductivity. A simple estimate
of the peak positions is the following. Scattering by surface inhomogeneities
changes the tangential momentum by $\Delta q\sim\pi/R$. This is sufficient for
the interband transition when this $\Delta q\sim q_{i}-q_{i+1}$. When the
number of occupied minibands is large, the lateral Fermi momentum for the
gliding electrons, \emph{i.e., }electrons from the miniband with a relatively
small index $i,$ $q_{i}\sim p_{F}$. For such electrons, $q_{i}^{2}-q_{i+1}%
^{2}\sim2\pi\Delta q/\lambda_{F}\sim2\pi^{2}/R\lambda_{F}$. On the other hand,
the energy conservation law dictates $q_{i}^{2}-q_{i+1}^{2}=\left(
2i+1\right)  \pi^{2}/L^{2}.$ Accordingly, with increasing $L$ the transition
channel $i\leftrightarrow i+1$ opens at $L^{2}\sim\left(  i+1/2\right)
R\lambda_{F}$. The opening of a new scattering channel in the points
\begin{equation}
L_{i}\sim\sqrt{\left(  i+1/2\right)  R\lambda_{F}} \label{b10}%
\end{equation}
is always accompanied by a drop in conductivity. The first such drop occurs
for the electrons in the lowest miniband $\epsilon_{1}\left(  \mathbf{q}%
\right)  $ with $i=1$, \emph{i.e.,} for the grazing electrons. Note, that
these particular electrons contribute the most to the conductivity. Since the
electrons from the lowest miniband are responsible for the dominant
contribution to the conductivity, the conductivity drops almost by half in the
point $L_{1}\sim\sqrt{3R\lambda_{F}/2}$ where $W_{12}$ becomes comparable to
$W_{11}$ and the effective cross-section doubles. [In the quasiclassical film
without bulk scattering, the current, which is an integral over momenta,
diverges when the component of momentum perpendicular to the film goes to
zero, \emph{i.e.}, for the grazing electrons. Without the bulk scattering, the
conductivity is finite only because of the quantum cut-off at $p_{x}=\pi/L$].

The anticipation was that this type of QSE should manifest itself also for the
interface scattering in multilayer films at $R\gg L$ for exponentially
decaying surface correlators. Indeed, such a picture can be observed in
Figures 3 and 4 for $u_{0}=0.1;10$ respectively (in both figures,
$L_{1}=1.1\lambda_{F},\ R=200\lambda_{F}$). The positions of the peaks in
Figure 3 for the weak interface are close to Eq. $\left(  \ref{b10}\right)  $.
In the case of the strong interface, the shift of the energy levels from those
for an "empty" square well is much more noticeable and the positions of the
peaks in Figure 4 deviate from those given by Eq. $\left(  \ref{b10}\right)
.$ At large values of $L$, the positions of the peak with strong interface
become close to the points in which the thickness of the second layer,
$L_{2}=L-L_{1}$, rather than the overall thickness $L$ is given by Eq.
$\left(  \ref{b10}\right)  $. The amplitude of the anomalous QSE oscillations
grows with the increasing strength of the interface approaching that for the
impenetrable wall.%

\begin{figure}[ptb]
\includegraphics{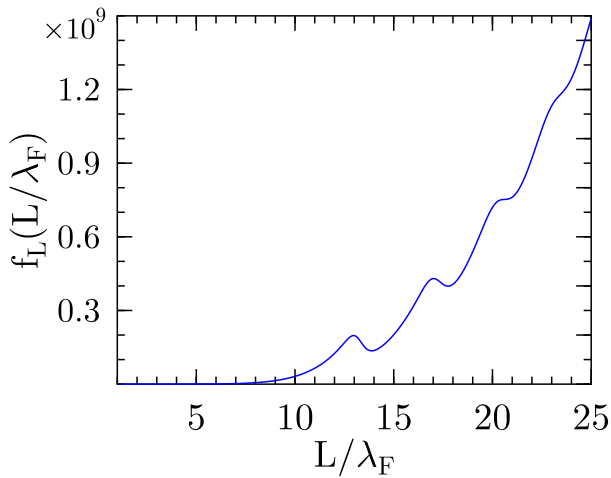}
\caption
{Anomalous QSE in conductivity of quantized films, Eq. (15), as a function of the film thickness $L$.
The correlation radius of inhomogeneities is large $R/ \lambda_F =200$,
the thickness of the first layer $L_1/ \lambda
_F =2.1$, the width of the interface
$d=0.1$, and the strength of the interface barrier $u_0=0.1$.}
\label{fig3}
\end{figure}%
%

\begin{figure}[ptb]
\includegraphics{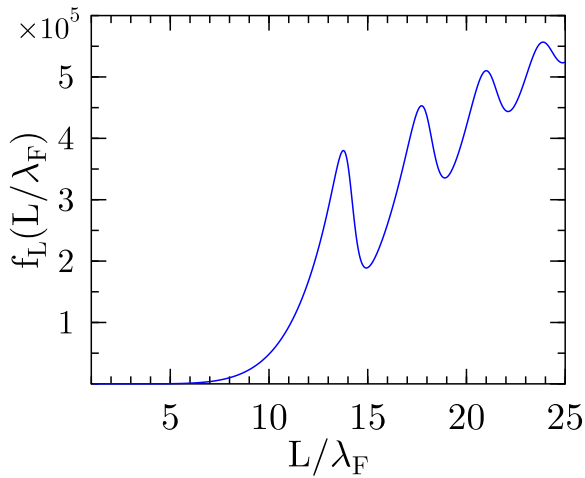}
\caption{The same as in Figure 3, but for a much stronger interface barrier,
$u_0=10$.}
\label{fig4}
\end{figure}%

Of course, for the inhomogeneities of the intermediate size, the picture
exhibits the features of both standard and anomalous QSE. As it has already
been mentioned, our numerical examples address the experiment in which the
size of the inhomogeneities $R$ is fixed while the thickness of the film
$L$\ is changing. In general, at the values $L<R$ one should see the smooth
anomalous QSE oscillations with large period, while at $L>R$ one should, on
the same curve, see the reappearance of the standard QSE with sharper
oscillations with period equal to 1. Roughly, the transitions between the
regimes occurs when the distance between the peaks of the anomalous QSE, Eq.
$\left(  \text{\ref{b10}}\right)  $, decreases to the value $\left(
L_{i+1}-L_{i}\right)  /\lambda_{F}\sim1$. In principle, the reappearance of
the standard QSE should be seen in Figures 3 and 4 when the computations are
extended to sufficiently large $L$. However, the amplitude of the standard QSE
oscillations on these curves is very small and the reappearance of the
oscillations is barely noticeable on the scale of the curve. It is much more
illustrative to demonstrate the effect at intermediate values of $R$ when both
anomalous and standard QSE oscillations have comparable amplitude. This is
shown in Figure 5 for $R/\lambda_{F}=3$ and weak interface $u_{0}=0.1$. On the
left side of the graph one can clearly see smooth "new" oscillations with a
relatively large period, while on the right side the oscillations recover the
sharp saw-like structure with the period equal to 1.%

\begin{figure}[ptb]
\includegraphics{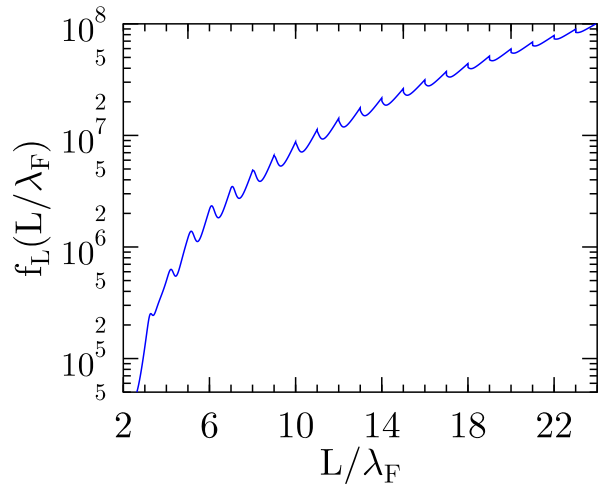}
\caption
{QSE in conductivity of quantized films, Eq. (15), as a function of the film thickness $L$ for the
intermediate values of the size of inhomogeneities, $R/ \lambda_F =3$.
The thickness of the first layer $L_1/ \lambda
_F =1.1$, the width of the interface
$d=0.1$, and the interface barrier $u_0=0.1$. At small $L$,
the curve exhibits the smooth oscillations
of the anomalous QSE with a large period, while QSE for large $L$ recovers the
standard saw-like shape with the period equal to 1. }
\label{fig4a}
\end{figure}%

\subsection{Geometric (fractional) quantum size effect}

To exhibit the QSE oscillations of the previous subsection, Figures 3 and 4
were plotted not for the exact $\delta-$type interfaces $\left(
\ref{b1}\right)  ,\left(  \ref{b2}\right)  $ but for a somewhat smeared (less
sharp) interface%
\begin{equation}
\delta U=-U_{0}\xi\left(  y,z\right)  \left[  \delta^{\prime}\left(
x-L_{1}\right)  +D\delta^{\prime\prime}\left(  x-L_{1}\right)  \right]  .
\label{b11}%
\end{equation}
The interface width $D$ can have two origins. If its origin is
corrugation-related, then the interface width is given by the next term of
expansion of the interface barrier in $\xi$ and is characterized by the same
parameters $\ell$ and $R$, $D^{2}\sim\left\langle \xi^{2}\right\rangle $. In
this case, depending on the correlation function, $D\sim\ell$ or $D\sim
\ell^{2}/R$. On the other hand, $D$\ can originate from some "internal"
smearing of the interface and can exist even without surface inhomogeneities.
In this case, $D$ is a new independent small parameter. Note, that here we are
interested in the "smearing" of the interface and not in its "fixed" width so
that the average of the square of the matrix elements of $\delta U$\ over the
interface starts from $D^{2}$. In Figures 3,4, the interface thickness was
chosen as $d=D/\lambda_{F}=0.1$.

If the interface is thinner, the character of the curves changes dramatically.
For example, Figure 6 presents the conductivity $\sigma\left(  L\right)  $
exactly for the same values of all parameters as in Figure 3 except for the
interface thickness which is now $d=D/\lambda_{F}=0.0001$. The difference
between the two curves is astonishing.%

\begin{figure}[ptb]
\includegraphics{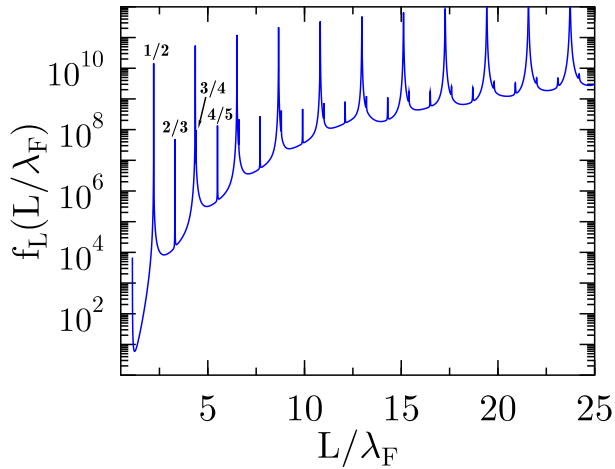}
\caption{Geometric QSE in conductivity of multilayer films.
The same parameters as in Figure 3, except for a much sharper interface,
$d=10^{-4}%
$. The fractions near the spikes give the values of the resonance positions
of the interface $\delta= L_2 /L$.}
\label{fig5}
\end{figure}
%

\begin{figure}[ptb]
\includegraphics{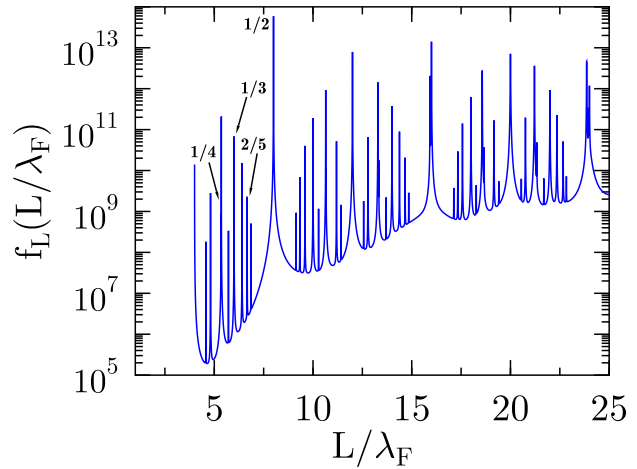}
\caption{Geometric QSE in conductivity of multilayer films.
The same parameters as in Figure 6, except for a wider first layer, $L_1 / \lambda
_F =4.0$ .
The fractions near the spikes give the values of the resonance positions
of the interface $\delta= L_2 /L$.}
\label{fig6}
\end{figure}%

The conductivity in Figure 6 exhibits two types of spikes. The explanation for
first type of spikes is the following. The scattering-driven transition
probabilities $W_{ij}$, Eq.$\left(  \ref{b4}\right)  ,$ contain the factor
with the correlation function $\zeta\left(  \mathbf{q}_{i}-\mathbf{q}%
_{j}^{\prime}\right)  $ and the geometric coefficients $G_{ij}$, Eq.$\left(
\ref{b5}\right)  .$ For exponentially decaying correlators with large $R\gg
L,$ the off-diagonal values of the correlation function $\zeta\left(
\mathbf{q}_{i}-\mathbf{q}_{j}^{\prime}\right)  $ with $i\neq j$ are
exponentially small in comparison with the diagonal ones, $\zeta\left(
\mathbf{q}_{i}-\mathbf{q}_{i}^{\prime}\right)  $. Then it is sufficient to
analyze only the diagonal elements of the matrix $G_{ij}$, Eq.$\left(
\ref{b5}\right)  :$
\begin{equation}
G_{ii}=4\Psi_{i}^{2}\left(  L_{1}\right)  \Psi_{i}^{\prime2}\left(
L_{1}\right)  . \label{b12}%
\end{equation}
If, accidentally, the $\delta-$type interface is positioned in the points in
which either $\Psi_{i}\left(  L_{1}\right)  =0\ $or $\Psi_{i}^{\prime}\left(
L_{1}\right)  =0$, then the coefficient $G_{ii}$, and, therefore, the
transition probability $W_{ii}$, become zero. This, in turn makes the
conductivity of electrons in the miniband $\epsilon_{i}$, and , therefore, the
overall conductivity, almost infinite. The cut-off is determined by one of
three factors: 1) exponentially small interband transitions; 2) scattering by
other defects such as impurities, inhomogeneities of external walls,
\emph{etc.}; 3) smearing of the interface $\left(  \ref{b11}\right)  $ that
leads to the averaging of $G_{ii},$ Eq. $\left(  \ref{b5}\right)  ,\left(
\ref{b12}\right)  $ over a finite interval making it non-zero. In this paper,
for obvious reasons, we are interested in the third option. Note, that in the
case of scattering by external film walls instead of the interlayer interface,
the coefficients $G_{ij}\sim i^{2}j^{2}$ are never equal to zero and this type
of QSE does not exist.

The first type of spikes corresponds to $\Psi_{i}\left(  L_{1}\right)  =0$.
The "resonance" positions of the $\delta-$type interface are universal and do
not depend on the potential strength. This is true for all rational points
$\delta=L_{2}/L$. Of course, the conductivity of the film becomes infinite for
this position of the interface \emph{only }\ if the corresponding miniband
$\epsilon_{i}$ is occupied. This means that the integer $n$ in the denominator
of the corresponding fraction $\delta=m/n$ should not exceed the number of the
occupied minibands, $n\leq S=\mathrm{Int}\left[  L/\lambda_{F}\right]  $.
Indeed, for points $\delta=L_{2}/L=m/n$ there is a number of the wave
functions $\Psi_{i}\left(  x\right)  $ of the \emph{empty} well that have
nodes in the points $x=L_{1}$. Since the unperturbed homogeneous potential
barrier has a $\delta-$functional form $U_{0}\delta\left(  x-L_{1}\right)  $,
these wave functions $\Psi_{i}\left(  x\right)  $ remain the eigenfunctions of
the well \emph{with} the unperturbed barrier $U_{0}\delta\left(
x-L_{1}\right)  $ inside and retain their nodes in the points $x=L_{1}$. Then
the corresponding diagonal coefficients $G_{ii}$ are zero making the diagonal
roughness-driven transition probabilities $W_{ii}$ for particles from the
miniband $\epsilon_{i}$ equal to zero as well. Since the off-diagonal
transition probabilities are exponentially small in $R/L\gg1$, the condition
$W_{ii}=0$ makes the conductivity for particles from the miniband
$\epsilon_{i},$ and, therefore, the overall conductivity exponentially large
in $R/L\gg1$.

The structure of the corresponding resonance spikes becomes more and more
complicated with an increase in $L_{1}$ when the structure of the minibands
and their occupancy become more convoluted. The simplest structure is observed
when $L_{1}$ is between $\lambda_{F}$ and $2\lambda_{F}$ as in Figure 6. In
this case, the observed rational spikes correspond to the rational numbers of
the form $\delta=\left(  n-1\right)  /n$ and are equidistant with the
separation $L_{1}/\lambda_{F}$. The first spike corresponds to the film with
$\delta=L_{2}/L=1/2$, the second - to $\delta=2/3$, the third - to
$\delta=3/4$, the fourth - to $\delta=4/5$, and so on. The odd peaks, with the
exception of the first one, look wider and consist of bigger and smaller
sub-peaks. The smaller sub-peaks correspond to the geometrical resonance with
$\delta=\left(  n-1\right)  /n$ which is described above. The bigger and wider
sub-peaks have a somewhat different nature and are not universal with respect
to the barrier strength. These sub-peaks will be described later. Note, that
the peak $\delta=3/4$\ is so close to the first peak from the other series
that these two peaks are hardly distinguishable.

When $L_{1}$ becomes bigger, the first few geometric resonances can be
observed at much narrower second layers, well before the point $\delta
=L_{2}/L=1/2$, while the density of the resonances become higher. For example,
Figure 7 presents the conductivity as a function of thickness for the film
with the same parameters as in Figure 6 except for the thickness of the first
layer which is now $L_{1}=4\lambda_{F}$. Though the overall distribution of
the peaks is now much more complicated, the majority can still be understood
as the ones generated by the eigenfunctions of the empty quantum well with the
nodes in the positions of the barrier. The complexity of the peak structure is
explained by the fact that at wider first layer $L_{1}$\ more minibands are
occupied thus allowing a wider selection of the rational numbers that
determine the peak positions $\delta=L_{2}/L=m/n$.

The geometric resonances can coexist with the anomalous QSE of the previous
section if the interface is relatively strong as in Figure 8 for the same
configuration as in Figure 6 but with much higher value of $u_{0}$, $u_{0}%
=10$. For weak interfaces, the geometric resonances suppress the QSE of the
previous subsection which gets restored only for bigger values of the
interface thickness $d$. This graduate disappearance of the geometric
resonances can be seen when comparing Figure 6 for $u_{0}=0.1,$ $d=10^{-4}$
with Figure 9 ($d=10^{-2}$) and Figure 10 ($d=10^{-1}$). Figure 10 presents
conductivity for the same configuration as Figure 3 but in logarithmic scale.
In this scale, one can see both the wide QSE oscillations of the previous
subsection and the only surviving geometric resonance at $\delta=1/2$.%

\begin{figure}
[ptb]
\includegraphics
{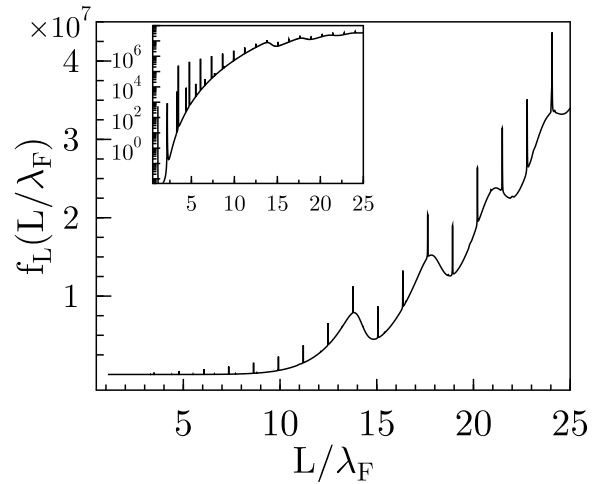}%
\caption{Geometric and anomalous QSE in conductivity of multilayer
films for strong interface potential, $u_0 =10$. The rest of the
parameters are the same parameters as in Figure 6. For easier
comparison with Figures 4 and 6, the insert gives the same data in
logarithmic scale.} \label{fig7}
\end{figure}

\begin{figure}[ptb]
\includegraphics{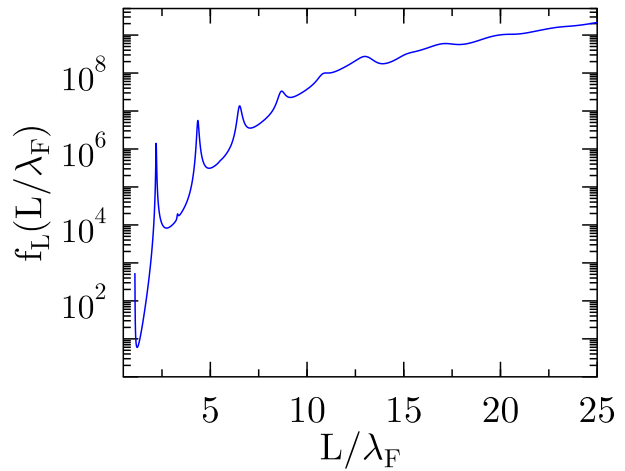}
\caption
{Illustration of the effect of smearing of the interface. The same curve as in Figure 6, but for a wider interface, $d=0.01$.}
\label{fig8}
\end{figure}%
%

\begin{figure}[ptb]
\includegraphics{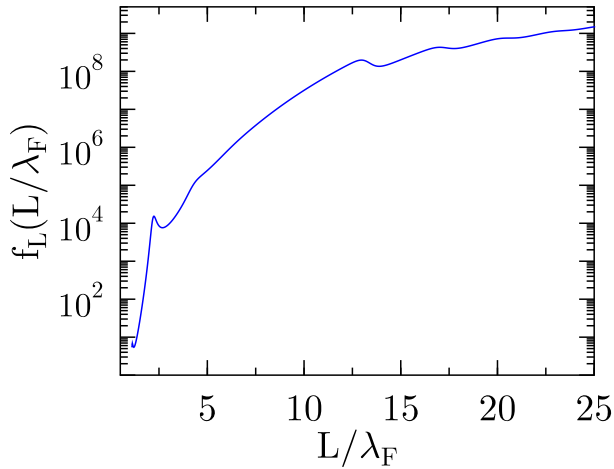}
\caption
{Illustration of the effect of smearing of the interface. The same curve as in Figures 6 and 9, but for an even wider interface, $d=0.1$.
Data as in Figure 3, but in logarithmic scale.}
\label{fig9}
\end{figure}%

Above we explained only\ the narrower, universal geometric resonances at
$\delta=m/n$ in Figures 6 - 9. The second, non-universal type of spikes has a
similar explanation - zeroes of $G_{ii}$. According to Appendix,%
\begin{align}
G_{ii}  &  =4\Psi_{i}^{2}\left(  L_{1}\right)  \Psi_{i}^{\prime2}\left(
L_{1}\right) \nonumber\\
&  =\frac{4\pi^{2}}{L^{4}}A_{i}^{4}\sin^{2}\left(  \pi z_{i}\delta\right)
\left[  g\sin\left(  \pi z_{i}\delta\right)  +2z_{i}\cos\left(  \pi
z_{i}\delta\right)  \right]  ^{2}, \label{b15}%
\end{align}
where $z_{i}\left(  g,\delta\right)  $ is given by the solution of the $1D$
Schroedinger equation $\left(  \ref{b9}\right)  $ for a quantum well with a
$\delta-$type barrier inside. The explicit form of the coefficients $A_{i}$ is
not important. The factor $\sin^{2}\left(  \pi z_{i}\delta\right)  $ in Eq.
$\left(  \ref{b15}\right)  $ corresponds to $\Psi_{i}^{2}\left(  L_{1}\right)
$; its zeroes are responsible for the geometric resonances with rational
$\delta=\left(  n-1\right)  /n$. There are no other zeroes of $\sin^{2}\left(
\pi z_{i}\delta\right)  $.

However, Eq. $\left(  \ref{b15}\right)  $\ also contains the factor in the
square brackets which corresponds to $\Psi_{i}^{\prime}\left(  L_{1}\right)
$. The simultaneous solution of Eq. $\left(  \ref{b9}\right)  $ for the
spectrum, $g\sin\left(  \pi z\delta\right)  +2z\cos\left(  \pi z\delta\right)
=0,$ and equation $\Psi_{i}^{\prime}\left(  L_{1}\right)  =0$\ yield the
following equation for the resonance positions of the interface:%
\begin{equation}
\sin\left[  \pi z_{i}\left(  1-2\delta\right)  \right]  =0, \label{b17}%
\end{equation}
which is equivalent to
\begin{equation}
z_{i}\left(  g,\delta\right)  \left(  1-2\delta\right)  =k \label{b18}%
\end{equation}
with integer $k$. First, there is a universal solution $\delta=1/2$ at $k=0$.
In this case, when the interface is exactly in the middle of the film, both
$\Psi_{i}\left(  L_{1}\right)  $ and $\Psi_{i}^{\prime}\left(  L_{1}\right)  $
are zero (the former with an even index, the latter - with the odd). This
explains why the geometric resonance with $\delta=1/2$\ is the most stable one
with respect to the smearing of the interface.

The rest of the resonances with $k\neq0$ are not universal. These resonances
explain the earlier unaccounted for spikes in Figures 6 - 8. Since the
spectrum $z_{i}\left(  g,\delta\right)  $ is a complicated function of the
interface strength and its position, the solution of Eq. $\left(
\ref{b18}\right)  $ for $k\neq0$ is rather complicated. We will give the
analytic equation for the simplest case of $\lambda_{F}\leq L_{1}<2\lambda
_{F}$\ when Eq. $\left(  \ref{b18}\right)  $ is equivalent to%
\begin{equation}
\frac{L}{\lambda_{F}}=\frac{2z_{n}}{z_{n}-n+1}\frac{L_{1}}{\lambda_{F}}
\label{b19}%
\end{equation}
(the only allowed values of $k$ are $k=-n+1$). For weak interfaces $g/z_{n}%
\ll1$ this equation can be rewritten as%
\begin{equation}
\frac{1}{1-\delta}=2n\left[  1-\frac{n-1}{n}\Delta_{n}\right]  ,\ \Delta
_{n}\approx\frac{g}{\pi z_{n}}\sin^{2}\left(  \pi n\delta\right)  .
\label{b20}%
\end{equation}
Analysis of Eq. $\left(  \ref{b20}\right)  $ shows that several first of such
resonance positions of the interface are indeed close to the odd rational
universal resonances as in Figure 6,%
\[
\delta\simeq\frac{2l-1}{2l},
\]
and separate from the universal resonances with increasing integer $l$. The
very first resonance at $\delta=1/2$ is, as it is explained above, exactly the
same as the first universal resonance. The reason why these non-universal
resonances are wider and stronger than the universal ones described above is
still unclear.

\section{Conclusions}

In summary, we analyzed QSE in conductivity of multilayer films when the main
scattering mechanism is the scattering of electrons by random inhomogeneities
of the interlayer interface. Three different types of QSE are predicted.

The first one is a standard QSE with a typical saw-like dependence of the
conductivity $\sigma$\ on the film thickness $L$, $\sigma\left(  L\right)  $.
This effect dominates when the correlation radius (size) of the interface
inhomogeneities $R$ is much smaller than the film thickness, $R\ll L$. This
effect should be observed for all types of the correlation functions of the
interface roughness. This effect is easily explained by the singularities in
the electron density of states related to the quantization of motion across
the film.

The second type of QSE is explained not by the quantization-driven
singularities in the density of states, but by the anomalies in the
cross-section for scattering by interface inhomogeneities. This
scattering-driven QSE replaces the standard saw-like QSE when the correlation
radius (size) of the interface inhomogeneities is large, $R\gg L$. This type
of QSE manifests itself as smooth large-scale oscillations on the dependence
$\sigma\left(  L\right)  $ and should be observed only when the Fourier image
of the interface correlation function (the so-called power spectrum of
inhomogeneities) decays exponentially at large momenta. The main difference of
this QSE from a similar effect in scattering by the film walls \cite{pon1} is
that the observation of this effect in multilayer film requires certain
smearing of the interface.

The third type of QSE is new and is most unusual. This effect manifests itself
as a set of very narrow and high spikes in $\sigma\left(  L\right)  $ and
replaces the scattering-induced QSE described above when the interface is
narrow. The finite cutoff in the spikes can be ensured either by some other
scattering mechanism or by the smearing of the interface.

The spikes are observed only for certain resonance positions of the interface.
The number of spikes is determined by the relation between the thickness of
the layers and the Fermi wavelength. The resonance positions of the interface
are described. These positions can be split into two general classes. Some of
these positions are universal and do not depend on the amplitude of the
interface potential barrier and correspond to the situations when the ratio of
the layer widths is given by simple rational fractions. The integer in the
denominator of such fractions does not exceed the number of occupied
minibands. The remaining resonance positions of the interface are
non-universal and depend on the strength of the interface potential. In the
case of weak interface, some of these non-universal positions are close to the
universal ones giving the impression of a split in the conductivity spikes.

Too small width of the resonance spikes can impede the experimental
observation of the geometric resonances. The width of the resonance spikes
increases and their height decreases with increasing smearing of the interface
and the resonance spikes gradually disappear. Note, that this disappearance of
the resonance spikes is related not to the widening, but to the random
smearing of the interface - the widening of the interface, by itself, results
just in a shift of the spike positions. The width of the universal resonances
is equal, by the order of magnitude, to the width of the smeared interface
$D$, Eq. (\ref{b11}), or, in dimensionless variables, $d=D/\lambda_{F}$. The
width of non-universal resonances is somewhat larger and is less sensitive to
$D$; the reason is still unclear. Note, that the smearing width $D$ can be
much smaller than the physical thickness of the interface, which in metals is
often larger than or of the order of the Fermi wavelength $\lambda_{F}$. In
contrast to this, the smearing parameter $d=D/\lambda_{F}$ can be very small.
The most stable spike with respect to smearing corresponds to the layers of
equal width. The wide range of possible values of $D$, which are determined
either by the roughness with $D$ of the order of $\ell$ or $\ell^{2}/R$ or by
the "internal" smearing of the interface, makes the observation of the
geometric resonances possible.

The spikes in the conductivity occur when the scattering probabilities for
electrons in one of the quantized minibands become exactly zero. Since
scattering probabilities for scattering by different interfaces add up, the
spikes in conductivity of multilayer films with many layers can be observed
only if the scattering probabilities for electrons from one miniband become
zero simultaneously for scattering by all the interfaces. This can happen only
if all the interfaces are located in the universal resonance positions
corresponding to the rational fractions from the same series. Otherwise, the
scattering by inhomogeneities of the "non-resonant interface" will curtail the
contributions from the resonant ones. This imposes a restriction on the number
of layers for an observation of this type of QSE for a film of fixed overall thickness.

We analyzed the multilayer films under the condition that the disruption in
the electron spectrum is caused only by the interface potential while electron
potential deep into the layers is the same for all layers. One can imagine a
different physical situation when the electron potential in different layers
differ from each other as in Ref. \cite{arm2}. In this situation the resonance
spikes in conductivity should be observed when the position of the interface
coincides with one of the nodes in the wave function. It is clear that this
occurs at least for certain values of the interlayer potential difference
$\Delta U$.

The calculations in the paper are aimed primarily at the experimental setup
when the lateral conductivity is measured as a function of the film thickness
at fixed thickness of the first layer (fixed position of the buried
interface). The main obstacle for the experimental observation of the
predicted effect is a rather small width of the conductivity spikes and their
sensitivity to the position of the interface. On the other hand, this
sensitivity of QSE to the position of the interface may open the door for
using this effect for precision control of the interface positions in
multilayer films. This may be very useful for better quality ultrathin films
without short-range surface inhomogeneities \cite{coupl1}. Recent experiments
with controlled ultrathin metal films with buried rough interfaces \cite{alt1}
indicate that the existing experimental setups are sufficient for the
observation of the predicted quantum size effect.

Usually, QSE in conductivity of semiconductor films is less pronounced than
for the metal films. This is explained by the smoother distribution of
electrons in non-degenerate semiconductors. In the absence of sharp drop in
the distribution at the Fermi energy, singular features in the conductivity,
which is an integral over the particle distribution, tend to be smeared out.
However, the universal geometric spikes in conductivity, which are described
above, are explained by the zeroes in quantized electron wave functions on the
interface and have nothing to do with the electron distribution. Then these
spikes in conductivity can be the only striking common feature for\ QSE in
multilayer metal and semiconductor films. The only obstacle for
observation\ of such spikes in semiconductors could be a relatively large
screening radius which may lead to an effective smearing of the interface.

This work was supported by NSF grant DMR-0077266.

\section{Appendix: Energy spectrum and matrix elements}

One-dimensional Schr\"{o}dinger equation for a square well with a $\delta
$-functional barrier inside has the form%
\begin{equation}
\psi\left(  x\right)  ^{\prime\prime}+k^{2}\psi\left(  x\right)  =u_{0}%
\delta\left(  x-a\right)  \psi\left(  x\right)  , \label{ap1}%
\end{equation}
where%
\begin{equation}
k^{2}=2mE/\hbar^{2},\ u_{0}=2mU_{0}/\hbar^{2}. \label{ap2}%
\end{equation}
The wave functions can be written as%
\begin{align}
\psi_{1}  &  \equiv\psi\left(  x\leq L_{1}\right)  =\sqrt{\frac{2}{L}}A\sin
kx,\label{ap3}\\
\psi_{2}  &  \equiv\psi\left(  x\geq L_{1}\right)  =\sqrt{\frac{2}{L}}B\sin
k\left(  x-L\right)  .\nonumber
\end{align}
\qquad

In dimensionless notations of Sec. II.2, the equation on spectrum acquires the
form%
\begin{equation}
\sin\left(  \pi z\right)  +\frac{g}{z}\sin\left(  \pi\delta z\right)
\sin\left[  \pi\left(  1-\delta\right)  z\right]  =0, \label{ap4}%
\end{equation}%
\[
\delta=L_{2}/L\leq1/2,\ kL=\pi z,\ g=u_{0}L/\pi.
\]

The normalized coefficients in the wave function $\left(  \ref{ap3}\right)  $
are equal to%

\begin{align}
A_{n}  &  =\frac{1}{\sqrt{\delta+\left(  1-\delta\right)  t_{n}^{2}+t_{n}%
\sin\left(  z_{n}\pi\right)  /z_{n}\pi}},\label{ap5}\\
B_{n}  &  =A_{n}t_{n},\nonumber
\end{align}
where
\begin{equation}
t_{n}=-\frac{\sin\pi\delta z_{n}}{\sin\left[  \pi\left(  1-\delta\right)
z_{n}\right]  }. \label{ap6}%
\end{equation}

The explicit expression for the spectrum $\left(  \ref{ap4}\right)  $\ can be
given in the limiting cases of weak and strong potential barriers. If the
barrier is weak, $g/z\ll1$, the spectrum is%
\begin{equation}
z_{n}=n+\Delta_{n},\ \Delta_{n}\approx\frac{g}{\pi n}\sin^{2}\left(  \pi
n\delta\right)  . \label{ap7}%
\end{equation}

In the opposite case of strong interface $g\rightarrow\infty$, the spectrum
decouples into two independent series of levels for each layer:%
\begin{equation}
z_{n_{1}}=n_{1}/\delta,\ z_{n_{2}}=n_{2}/\left(  1-\delta\right)  .
\label{ap8}%
\end{equation}
For large, but finite $g$, the corrections to the spectrum $\left(
\ref{ap8}\right)  $ can be easily obtained by expansion in $z_{n}/g\delta$ or
$z_{n}/g\left(  1-\delta\right)  $:%
\begin{equation}
z_{n_{1}}\approx\frac{n_{1}}{\delta}\left(  1-\frac{1}{g}\right)  ,\ z_{n_{2}%
}\approx\frac{n_{2}}{1-\delta}\left(  1-\frac{1}{g}\right)  . \label{ap9}%
\end{equation}
An important restriction for Eq. $\left(  \ref{ap9}\right)  $ is that the
energy levels in each of the layers, Eq. $\left(  \ref{ap8}\right)  $, should
not be very close to each other. In the case of near degeneracy, the two close
levels, as usual, repel each other with a resulting gap equal to%
\begin{equation}
\Delta z_{n}\simeq\frac{z_{n}}{\pi g\delta\left(  1-\delta\right)  }.
\label{ap10}%
\end{equation}
The above equations should be modified if the interface is very close to one
of the external walls of the well, \emph{\ i.e.}, if either $\delta\ll1$ or
$1-\delta\ll1$.

Note, that if the $\delta$-type barrier is located exactly in the node of one
the wave functions of the empty well, this wave function remains the
eigenfunction of the well with a barrier inside irrespective of the strength
of the barrier. This means that the energy levels that correspond to such wave
functions are not shifted by the presence of the barrier.

The matrix elements of the roughness-related perturbation $\left(
\ref{b5}\right)  $ can be calculated with the help of the above functions
$\left(  \ref{ap3}\right)  ,\left(  \ref{ap5}\right)  $:%
\begin{align}
G_{nm}  &  =\left[  \Psi_{m}\left(  L_{1}\right)  \Psi_{n}^{\prime}\left(
L_{1}\right)  +\Psi_{m}^{\prime}\left(  L_{1}\right)  \Psi_{n}\left(
L_{1}\right)  \right]  ^{2}\nonumber\\
& =\left(  \frac{2\pi}{L^{2}}A_{m}A_{n}\right)  ^{2}g_{nm}^{2},\label{ap11}\\
g_{nm}  &  =g\sin\left(  \pi z_{m}\delta\right)  \sin\left(  \pi z_{n}%
\delta\right)  +z_{m}\cos\left(  \pi z_{m}\delta\right)  \sin\left(  \pi
z_{n}\delta\right) \nonumber\\
&  +z_{n}\cos\left(  \pi z_{n}\delta\right)  \sin\left(  \pi z_{m}%
\delta\right)  .\nonumber
\end{align}
The most important are the diagonal matrix elements%
\begin{align}
g_{nn}  &  =\sin\left(  \pi z_{n}\delta\right)  \left[  g\sin\left(  \pi
z_{n}\delta\right)  +2z_{n}\cos\left(  \pi z_{n}\delta\right)  \right]
\label{ap12}\\
&  =\frac{z_{n}\sin\left(  \pi z_{n}\delta\right)  \sin\left[  \pi
z_{n}\left(  1-2\delta\right)  \right]  }{\sin\left[  \pi z_{n}\left(
1-\delta\right)  \right]  }.\nonumber
\end{align}
Note that the zeroes of the denominator in Eq. $\left(  \ref{ap12}\right)  $
are cancelled out by the zeroes of $A_{n}^{2}$, Eq. $\left(  \ref{ap11}%
\right)  $. When the interface has a finite width $d$, the matrix elements
acquire the following addition:%
\[
g_{nm}^{2\left(  tot\right)  }=g_{nm}^{2}+d^{2}\left(  \Delta g_{nm}\right)
^{2}%
\]
where
\begin{align}
\Delta g_{nm} & =2z_{n}z_{m}\cos\left(  \pi z_{n}\delta\right)  \cos\left(
\pi z_{m}\delta\right) \nonumber\\
& -\left(  z_{n}^{2}+z_{m}^{2}\right)  \sin\left(  \pi z_{n}\delta\right)
\sin\left(  \pi z_{m}\delta\right)  +g\cdot g_{nm}..
\end{align}
The total matrix element, $G_{nm}^{\left(  tot\right)  }$, is never zero. This
means that the term with $d^{2}$, which originates from the smearing of the
interface, provides a natural cutoff for the conductivity.

In degenerate metal films, of all the energy minibands $\epsilon_{n}$ only the
minibands with $n\leq\mathrm{Int}\left[  L/\lambda_{F}\right]  $ are occupied.

\end{document}